\documentclass[reprint,preprintnumbers,superscriptaddress,aps]{revtex4-1}

\usepackage[utf8]{inputenc}
\usepackage{microtype}
\usepackage{graphicx}
\usepackage{mhchem}
\usepackage{siunitx}
\usepackage{booktabs}
\usepackage{xcolor}
\usepackage{epsf,color,soul,amsmath,amssymb}
\usepackage[dvips]{epsfig}
\usepackage{ulem}
\usepackage[T1]{fontenc}
\usepackage{lmodern}
\usepackage[colorlinks=true,linkcolor=blue,urlcolor=blue,citecolor=blue]{hyperref}
\usepackage[]{hyperref}

\newcommand\rsout{\bgroup\markoverwith{\textcolor{red}{\rule[0.5ex]{2pt}{0.4pt}}}\ULon}

\begin{document}

\author{M. Clementi}\email{ marco.clementi01@universitadipavia.it}
\affiliation{Dipartimento di Fisica, Università degli Studi di Pavia, Via Bassi 6, 27100 Pavia, Italy}
\author{A. Barone}
\affiliation{Dipartimento di Fisica, Università degli Studi di Pavia, Via Bassi 6, 27100 Pavia, Italy}
\author{T. Fromherz}
\affiliation{Institute of Semiconductor and Solid State Physics, Johannes Kepler University, Altenberger Str. 69, 4040 Linz, Austria}
\author{D. Gerace}
\affiliation{Dipartimento di Fisica, Università degli Studi di Pavia, Via Bassi 6, 27100 Pavia, Italy}
\author{M. Galli}
\affiliation{Dipartimento di Fisica, Università degli Studi di Pavia, Via Bassi 6, 27100 Pavia, Italy}

\title{Selective tuning of optical modes in a silicon comb-like photonic crystal cavity }

\begin{abstract}
Realizing multiply resonant photonic crystal cavities with large free spectral range is key to achieve integrated devices with highly efficient nonlinear response, such as frequency conversion, four-wave mixing, and parametric oscillation. This task is typically difficult owing to the cavity modes' sensitivity to fabrication disorder, which makes it hard to reliably achieve a comb-like spectrum of equally spaced modes even when a perfect matching is theoretically predicted. Here we show that a comb-like spectrum of up to 8 modes with very high quality factor and diffraction limited volumes can be engineered in the bichromatic-type potential of a two-dimensional photonic crystal cavity fabricated in a thin silicon membrane. To cope with the tight tolerance in terms of frequency spacings and resonance linewidths, we develop a permanent post-processing technique that allows the selective tuning of individual confined modes, thus achieving an almost perfect frequency matching of high Q resonances with record finesse in silicon microresonators. Our experimental results are extremely promising in view of ultra-low power nonlinear photonics in silicon.
\end{abstract}


\maketitle
% INTRODUCTION
\section{Introduction}
In the last few decades, photonic crystal (PhC) cavities attracted a considerable interest in the field of integrated optics thanks to their capability to effectively confine light, both in time and space \cite{Notomi2010}. The localization in time (spectrum) is quantified by the quality (Q) factor, ranging from few thousands \cite{Akahane2003} up to several millions \cite{Asano2017} for these  structures, while the spatial confinement is quantified by the mode volume ($V$) \footnote{Throughout the article we will always use the standard cQED definition of mode volume $V=\frac{\int{\varepsilon\left|E \right|^2 d^3r }}{\max\{\varepsilon\left|E \right|^2\}}$. }, which is typically of the order of $\left(\lambda/n\right)^3$ ($n$ being material refractive index). 
Due to these unique properties, PhC cavities are among the best suited devices for integrated optics application requiring strong light-matter interaction, which is enhanced with respect to the bulk case by the Q$/V$ figure of merit. The latter can reach its highest values for PhC cavities \cite{Asano2018}. 

In parallel, multi-modal optical microcavities exhibiting a comb-like spectrum have witnessed a strong attention in recent years thanks to their suitability for applications such as metrology \cite{Udem2002, Kippenberg2011}, optical parametric oscillators (OPO) \cite{Kippenberg2004,Razzari2010,Gaeta2019} and integrated sources of non-classical light based on spontaneous four-wave mixing (FWM) \cite{Harris2014,Reimer2016}. These structures are commonly based on whispering gallery mode resonators, such as microspheres, microtoroids and microrings, where equally spaced modes in energy naturally emerge. However, these implementations suffer from fairly large modal volumes, reducing the overall figure of merit Q$/V$ of the cavity and thus the efficiency of the nonlinear processes of interest \cite{Helt2012a}.

In this framework, it is desirable to implement microresonators exhibiting comb-like spectra (i.e. resonant modes equally spaced in energy) in PhC cavities, where the confinement mechanism yields diffraction-limited values of $V$. This kind of structures, however, does not naturally exhibit equally spaced modes, as they typically originate from point defects in an otherwise ideal two-dimensional PhC lattice. On the other hand, by coupling multiple identical PhC resonators in a \textit{photonic molecule}, the emergence of supermodes provides a way to achieve equally spaced resonances \cite{Cluzel2011,Azzini2013a}, but this approach suffers from larger mode volumes, poor spatial overlap between supermodes and a somewhat complex fabrication procedure. These issues can be overcome by exploiting a single, appropriately engineered multi-modal cavity. Recently, a PhC cavity design based on the superposition of two slightly mismatched lattices (from now on referred as \textit{bichromatic cavity}) was proposed by Alpeggiani \textit{et al.}  \cite{Alpeggiani2015}. It was shown, both theoretically and experimentally \cite{Combrie2017}, how this geometry provides a confinement potential for the electromagnetic field that effectively mimics the case of a particle in a harmonic potential.  This results, for an appropriate choice of design parameters, in nearly-equally spaced modes and Gauss-Hermite field envelope profiles. Nevertheless, even in these systems, the experimental realization of a multimodal PhC cavity with high-Q and equally spaced resonances remains hindered by the unavoidable presence of fabrication disorder, which leads to a statistical deviation of the resonant frequencies from perfect spacing by several linewidths \cite{Combrie2017}.    

In this work, we report the design, fabrication and characterization of PhC cavities based on a bichromatic potential and patterned in a suspended silicon membrane, engineered to exhibit comb-like resonant modes with measured Q factors exceeding 1 million in the telecom wavelegth range ($\lambda\sim1.55$~$\mu$m). 
We discuss 
the structural parameters exploited to theoretically optimize these devices to display at least three modes with equally spaced frequencies. 
We illustrate our experimental results obtained from the characterization performed by resonant scattering (RS) technique. 
As the main result of this work, we show how small mismatches due to the fabrication process can be compensated via permanent post-fabrication tuning performed by laser-induced local oxidation of the suspended membrane in order to achieve modes equally spaced within their linewidths, thus enabling the implementation of triply resonant nonlinear processes.

\begin{figure}[t]
\centering
\includegraphics[width=0.40\textwidth]{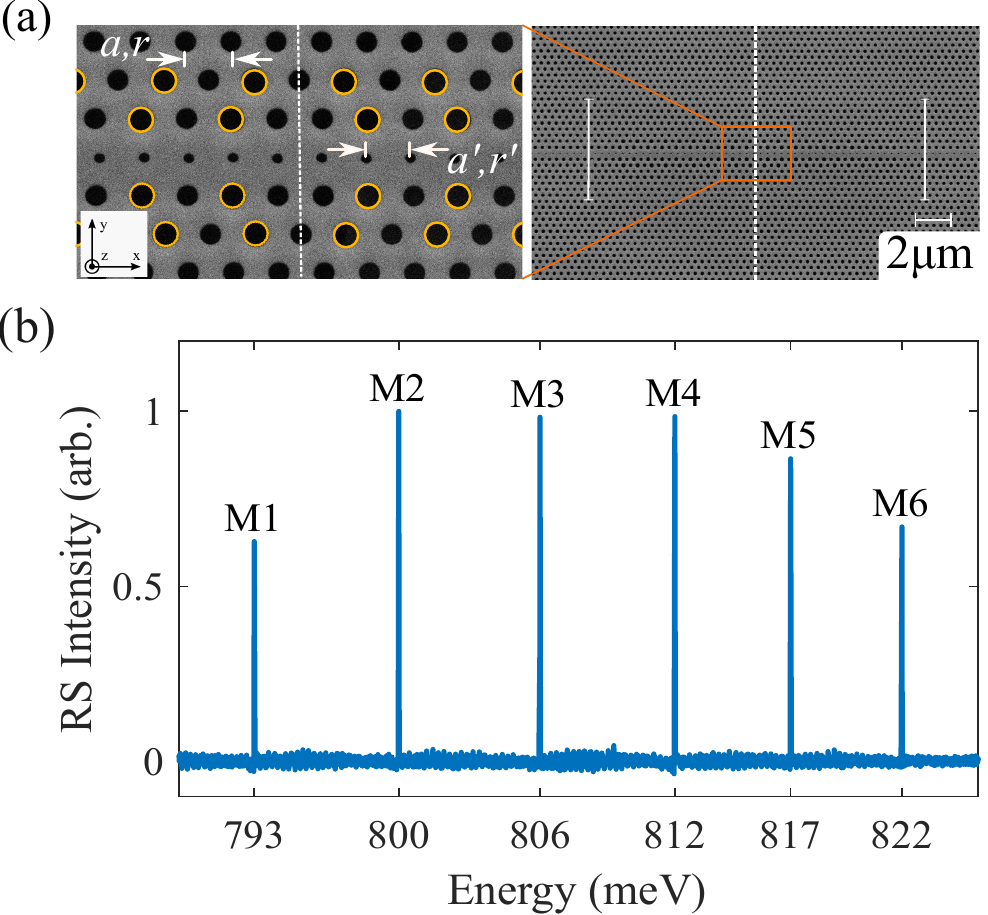}
\caption{(a) SEM image of one of the fabricated cavities. The two superimposed PhC lattices are characterized by periodicities $a$ and $a'$ and hole radii $r$ and $r'$ (see appendix \ref{sec:design_app} for details on cavity design). 
Holes marked by yellow circles have a radius increased by $\Delta r$ for far-field optimization (see, e.g., Refs.~\cite{Portalupi2010a,Tran2009}). In the right panel, the defect region is enclosed between the vertical white lines ($N=48$ defect holes with radius $r'$ and separated by $a'$), while in the remaining parts of the line defect, the original lattice constant $a$ is restored. 
(b) Measured broadband spectrum of one of the fabricated samples by Resonant Scattering (see main text). 
}\label{fig:1}
\end{figure}

The cavity designs employed in this work have been introduced and experimentally realized in previous works, with a focus on optimizing the fundamental mode Q-factor \cite{Alpeggiani2015,Simbula2016}. Briefly, light confinement in these structures stems from the superposition of two slightly mismatched preiodicities within a line-defect in the PhC lattice, as it is schematically shown in Fig.~\ref{fig:1} on a scanning electron microscope (SEM) image of one of the fabricated devices, in which the relevant parameters of the model are defined. Notably, the effective confining potential exhibits an approximately parabolic shape along the waveguide axis, which in turn results in equally spaced modes, as already shown for III-V based devices \cite{Combrie2017}. For the cavities employed in this work, $N=48$ defect holes spaced by $a'$ constitute the cavity potential, as detailed in appendix \ref{sec:design_app}, which sets the lattice mismatch to $a'/a = N/(N+1)\sim 0.98$ \cite{Simbula2016}. From full three-dimensional finite difference time domain (3D-FDTD) simulations \footnote{All FDTD simulations were performed with a commercial-grade software,\textit{ Lumerical Solutions, Inc}.}, we expected multi-modal spectra exhibiting more than 8 resonant modes, with an average free spectral range (FSR) of approximately $4$~meV, Q factors in excess of $10^7$, and  mode volumes of the order of $V=2\left(\lambda/n\right)^3$.\\
The devices were fabricated on a $220$~nm thick silicon-on-insulator (SOI) commercially available from SOITEC with $3$~$\mu$m thick buried oxide. The PhC pattern was defined by electron beam lithography (EBL) on a PMMA resist and successively transferred to the device layer by inductively coupled plasma reactive ion etching (ICP-RIE) based on low-pressure SF$_6$/O$_2$ mixture at cryogenic temperature ($-90~^\circ$C). Under-etching of the buried oxide was performed using hydrofluoric acid (HF) in 10\% aqueous solution, in order to obtain suspended membranes, and a subsequent dip in diluted HF (1\% aqueous solution), in order to reduce the overall roughness of the silicon/air interfaces. Further details about the fabrication process can be found in \cite{Simbula2016}. 
Quantitative analysis of the SEM images such as the one in Fig.~\ref{fig:1}a highlighted a good reproducibility in terms of holes' radii ($\sigma_r<2$~nm on the same chip) and lattice constant ($\sigma_a<1$~nm), where $\sigma$ denotes the standard deviation. \\
A systematic characterization of the fabricated devices was performed by the resonant scattering (RS) technique \cite{Galli2009}, as detailed in appendix \ref{sec:rs_technique}. In particular, we employed an optimized acquisition technique allowing to obtain single-resonance spectra with an overall absolute accuracy of $0.2~\mu$eV, which is below the minimum measured linewidth ($\Gamma_{min}\approx 0.7 ~\mu$eV). 
The measured RS spectra show up to 8 resonant modes, in agreement with simulations, with average FSR ranging from $\mathrm{FSR}_{5-6}=4.6$~meV for higher order modes to $\mathrm{FSR}_{1-2}=13$~meV. A typical broad band spectrum displaying a comb-like distribution with 6 resonant modes is shown in Fig.~\ref{fig:1}b. 
 
\begin{figure}[t]
\centering
\includegraphics[width=0.4\textwidth]{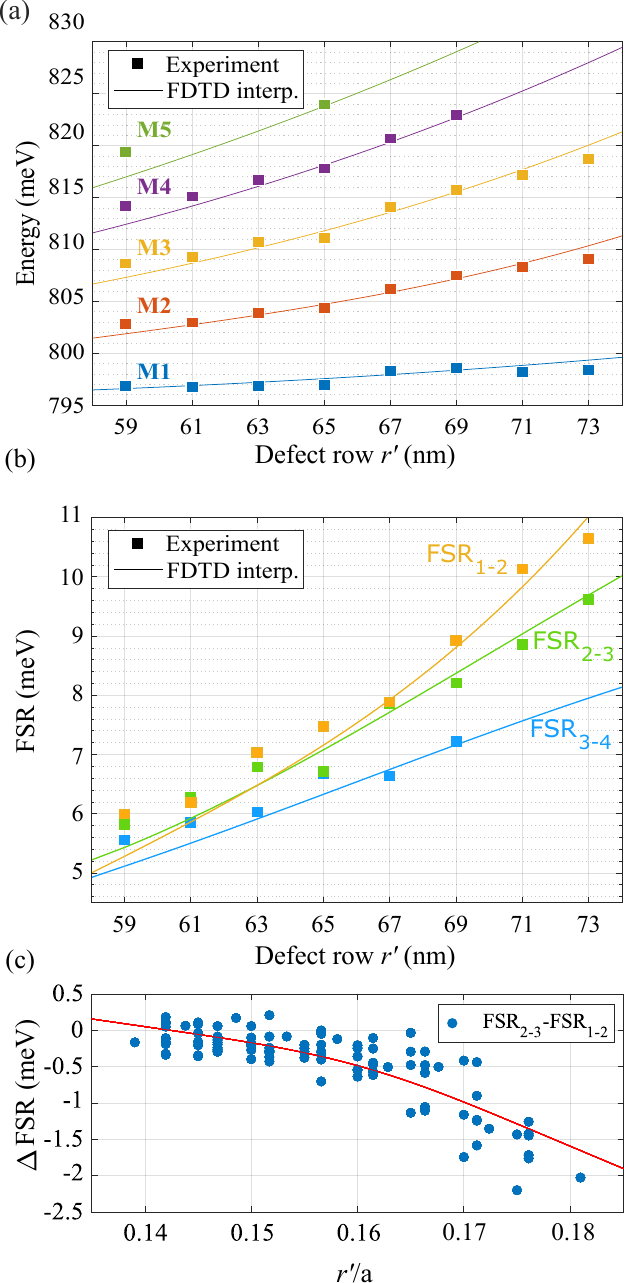}
\caption{(a) Resonance energy as a function of the line defect radii, $r'$, for the first 5 modes. Solid lines represent a polynomial interpolation of the FDTD data (calculated for $r'$ from 58 nm to 74 nm in steps of 2 nm), while squares represent the RS measured values.  (b) Free spectral range as a function of $r'$, for the first 4 modes. Again, solid lines represent a polynomial interpolation of the FDTD data, while squares represent the RS measured values. 
(c) Statistics of the FSR mismatch $\Delta \mathrm{FSR}_{i,j,k}=\mathrm{FSR}_{j-k}-\mathrm{FSR}_{i-j}$ for the first three resonances as a function of the $r'/a$ ratio over the entire poll of 106 measured devices. The red line is a guide to the eye. }\label{fig:2}
\end{figure}

Since the confining potential is only approximately harmonic, we introduced the inner hole radius, $r'$, as a scanning parameter to achieve the frequency matching condition for at least three resonant modes. In fact, by varying such parameters in the 3D-FDTD simulations we confirm that the resonance tuning depends on the mode order:  higher values of $r'$ yield an increased mode spacing, with a more pronounced effect appearing on the lower order modes, as shown in Fig.~\ref{fig:2}a, where full lines represent an interpolation to the simulated data. Remarkably, for low values of $r'$ it is possible to find a region where the trend is inverted and the lower order modes present a smaller FSR than the higher order ones, or even where two adjacent FSRs are matched within the modes' linewidths (see Fig.~\ref{fig:2}b). The latter condition would be highly desirable, as it would constitute the main requirement for a demonstration of triply resonant nonlinear processes, with huge potential for exploitation in FWM experiments.

% DESIGN AND FABRICATION
\section{Fabrication and experimental results}

To experimentally confirm our design strategy, we realized samples with lattice step of the main PhC varying from $a=400$~nm to $a=430$~nm, in order to achieve a modulation in the resonance wavelength via lithographic tuning, and the radius of the inner holes was nominally varied from $r'=59$~nm to $r'=73$~nm in steps of $2$~nm to tune the spacing of the comb-like spectrum of modes. Furthermore, the radius of specific holes (marked in yellow in Fig.~\ref{fig:1}a) was systematically modified in the range $\Delta r = \pm 4$~nm, in order to improve the coupling to free space in the direction perpendicular to the membrane, at the expense of a deliberate increase of the out-of-plane losses  \cite{Portalupi2010a, Tran2009}. The effect of such an optimization of the far-field pattern is twofold: on one side, it allows increasing the coupling efficiency of all the modes under vertical excitation from free-space, on the other it decreases the overall Q factor of the cavity (qualitatively, by introducing an extrinsic loss channel) and thus relaxing the condition of equally spaced modes (within the accuracy of a resonance linewidth). Remarkably, both effects were observed not only on the fundamental mode, but also on higher order ones.

\begin{figure*}[t]
\centering
\includegraphics[width=0.78\textwidth]{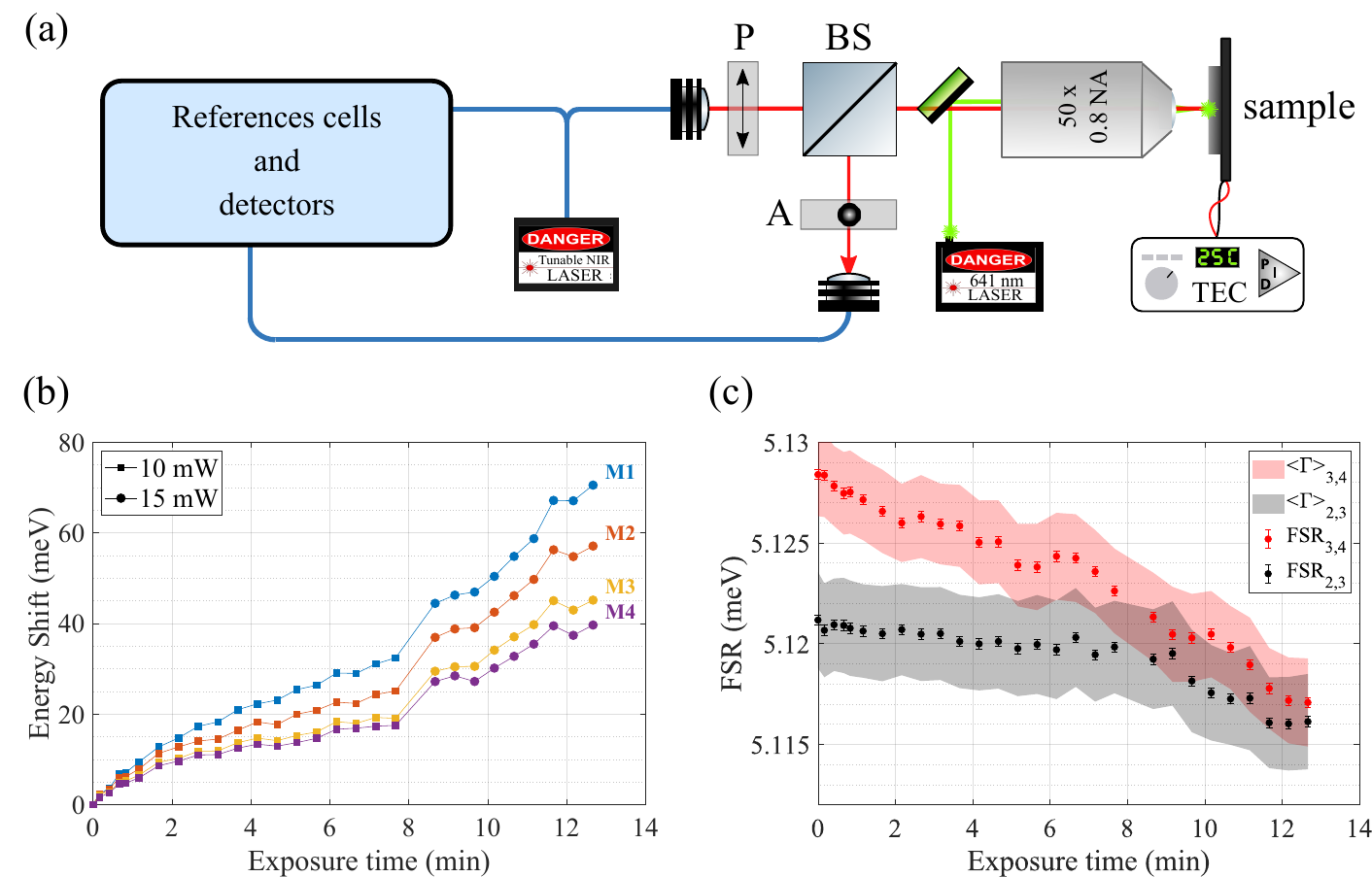}
\caption{(a) Schematic of the experimental apparatus (see appendix \ref{sec:rs_technique} for details). 
Notice that the same RS apparatus used for the sample characterization is exploited for the tuning. 
(b) Energy shift of the first 4 modes after local oxidation operated at the center of the PhC cavity. The vertical dashed line marks the increase of the oxidation power from 10~mW to 15~mW. The total shift depends on the mode order, with a more pronounced effect on lower order modes and a maximum overall shift of $69~\mu$eV for the fundamental. (c) FSR shifts of the pairs of modes 2,3 and 3,4 exploited in the tuning experiment. The average linewidths (full width half maximum) of the pairs of modes involved are represented as shaded areas. Error bars, which are specifically evaluated for each tuning step, are systematically smaller than mode linewidths.
}\label{fig:4}
\end{figure*}

% OPTICAL MEASUREMENTS

A comparison between the experimentally extracted resonant frequencies and the 3D-FDTD simulated ones  is reported in Fig.~\ref{fig:2}a as a function of $r'$, showing a very good overall agreement. From a  best fit of the measured resonances, we estimated Q factors as high as $1.1 \times 10^6$ 
 for cavities without far-field optimization. In fact, {the highest Q-factor was measured for the fundamental resonant mode of a cavity with $a=410$~nm, $r'=59$~nm, and resonant energy  $E_{\mathrm{M1}}=794.4$~meV, with $\mathrm{FSR}_{1-2}=5.65$~meV}. This corresponds to a finesse $\mathcal{F}= \mathrm{FSR}/\Gamma =8 \times 10^3$, {which is the highest reported value so far for a silicon microresonator,  to the best of our knowledge}, and comparable to the one of the state-of-art ring resonators based on silicon nitride \cite{Ji2017}.
The discrepancy between the experimental Q-factors and the FDTD prediction is attributed to fabrication imperfections and induced absorption at the etched interfaces \cite{Asano2017}, thus setting a practical limitation to technological applications of these devices. 
The Q-factor decreases to an average value around 240,000 
for far-field optimized cavities with $\Delta r = +4$~nm, which exhibit the best visibility observed in this work.

The measured trends in Fig.~\ref{fig:2}a follow the theoretical behavior, thus suggesting the possibility to find a combination of parameters for which the spacing between pairs of resonances is matched within the mode linewidth. However, even for far-field optimized cavities the intrinsically high Q factor of this bichromatic cavity design practically limits the applicability of this approach, as already discussed in Ref.~\cite{Combrie2017} for InGaP PhC cavities. Indeed, even in the presence of a fine lithographic tuning of the parameters, as it was performed in the present work, the accuracy required to the fabrication process to define the mode resonance energy is on the order of $\pm 5~\mu$eV for a far-field optimized cavity with a Q factor of ~240,000, i.e. unaccessible with the current technology. 
To give a quantitative picture of this statement, we report in Fig.~\ref{fig:2}b the FSR of the first few pairs of resonances as a function of $r'$, corresponding to the measured and simulated values of Fig.~\ref{fig:2}a. While the matching condition is theoretically predicted to occur at specific values of $r'$, this is hardly met in practice. In fact, we also report in Fig.~\ref{fig:2}c the accumulated statistics of the FSR mismatch from the first three modes on all the 106 devices measured in this work, as a function of the $r'/a$ ratio. The statistical deviation of the measured values is below $1\%$ of the resonance energy on the design parameters, which is comparable to the statistical error in the EBL fabrication as estimated from SEM imaging (remarkably, yet of the order of 1~nm). While the values accumulate around the desired zero FSR mismatch at lower $r'$ values, it is statistically very unlikely to actually find a device with the required characteristics.

In the light of these results, two possible strategies could be applied: either deliberately increasing the cavity modes linewidth by introducing a loading channel, such as an even heavier far-field optimization or the side-coupling to a waveguide, at the cost of a reduced efficiency for any nonlinear process, or introducing an \textit{a posteriori} selective tuning of one or more resonant modes with respect to the others, in order to compensate for the FSR mismatch. Here we propose a combination of these two options.
First, a cavity with FSR mismatch of only $\Delta\mathrm{FSR}=\mathrm{FSR}_{3-4}-\mathrm{FSR}_{2-3}=6~\mu$eV was selected among the measured devices. The far-field optimization of such device, obtained from a modification of the holes' radii $\Delta r = +4$~nm, corresponds to a measured Q factor of $~190,000$ on the three modes involved. By exploiting the same apparatus used for the RS measurements, we employed a focused visible laser (Coherent Cube, $\lambda_{ox}=641$~nm) to heat the device in order to locally oxidize the Si membrane as suggested by Chen \textit{et al.}  \cite{Chen2011}. A scheme of the whole setup is shown in Fig.~\ref{fig:4}a. The overall effect of this process on the first 4 cavity modes, as a function of the exposure time, is reported in Fig.~\ref{fig:4}b, where the local oxidation was induced by focusing the spot of the visible laser at the center of the cavity with diffraction-limited precision. A systematic blue-shift of all the modes involved is observed, pointing to an overall reduction of the average effective index in the spatial volume where the modes are localized. As a further aspect compared to Ref.~ \cite{Chen2011}, we notice here that the fundamental mode is mostly affected by the local oxidation, while higher order modes are systematically less affected by the process. 
This evidence is consistent with the calculated mode profile (see appendix \ref{sec:design_app}),
where the mode energy is mainly located in lobes of increasing distance from the cavity center, and can thus be interpreted as a first order perturbation emerging from the local variation of the refractive index $\Delta \lambda_i/\lambda_i \propto \Delta n/n$, as suggested in Refs.~\cite{Hennessy2006,Sokolov2015a}. \\
Finally, the effect of selective tuning on the FSRs is illustrated in Fig.~\ref{fig:4}c, where the average linewidth of the two modes is visualized as a shaded area, in order to highlight the accuracy to which the equal-spacing condition has to be fulfilled practically. As shown in Fig.~\ref{fig:4}c,  this condition is evidently achieved for the longest exposure times used in this work ($13$~min, 8 of which at a laser power of $10$~mW and 5~min at $15$~mW). Remarkably, the process results in an average overall shift of $53~\mu$eV while the FSR discrepancy compensated here amounts to only $6~\mu$eV. Finally, the regularity of the trends shown in figs.~\ref{fig:4}b,c points to the high degree of accuracy achievable with this tuning process, which can be further adjusted by a careful choice of the exposure time.

% CONCLUSIONS
\section{Conclusions}
In summary, we have presented the design and demonstration of a PhC cavity based on a bichromatic lattice exhibiting comb-like spectrum and best measured Q-factors above 1 million. The essential limitations to employing these devices emerge from the accuracy of the fabrication process, which limits the maximum value of Q achievable and the repeatability in the alignment of the resonant modes. As supported by recent experimental studies \cite{Asano2017}, we believe that the former issue is intimately connected with the standard fabrication process of silicon PhC devices, and in particular with the parasitic absorption provided by defect states at the interfaces. Conversely, we showed how the latter issue, which emerges as a consequence of the absence of a natural condition for the equal spacing of the resonances, can be overcome by post-fabrication tuning of the fabricated devices performed via local oxidation, which is a key result achieved in the present work. \\
In the perspective of nonlinear applications such as fully resonant frequency conversion, OPO and spontaneous or stimulated FWM, our devices and permanent post-fabrication tuning technique are extremely relevant owing to the silicon photonics compatible platform, to the large Q/V figure of merit attainable, which yields record-high values of finesse, and to the well known strong nonlinear response of silicon, all key-features for the implementation of low power nonlinear photonic devices.

\section*{Acknowledgments}
The authors would like to thank Alma Halilovic for significant support during the fabrication steps and Luca Zagaglia for participating in the initial modeling of far-field optimized bichromatic cavities.  This work was supported by the EU H2020 QuantERA ERA-NET Co-fund in Quantum Technologies project CUSPIDOR, co-funded by the Italian Ministry of Education, University and Research (MIUR) and by MIUR: “Dipartimenti di Eccellenza Program (2018-2022)”, Department of Physics, University of Pavia. Additional support by the Austrian Science Fund FWF under project number I3760 is acknowledged. We acknowledge support by COST Action MP1403 “Nanoscale Quantum Optics” through the Short Term Scientific Mission (STSM) program.

\appendix
\section{Bichromatic photonic crystal cavity: design and parameters }
\label{sec:design_app}
We hereby review the essential properties of our cavity design, based on previously published works \cite{Alpeggiani2015,Simbula2016}, for completeness.
The physics of light confinement in the \textit{bichromatic cavity} stems from the superposition of two slightly mismatched dielectric lattices, a condition originally studied in the framework of solid state physics and described by the so-called Aubry-André-Harper (AAH) model \cite{Harper1955,Aubry1980,Modugno2009}, which predicts the exponential localization of an electronic state whenever the mismatch $\beta=a'/a$ between the two lattice constants approaches an irrational value. 
As detailed in ref.~\cite{Alpeggiani2015}, the AAH model can be recast to the case of light propagation in dielectric systems, where the steady state Maxwell's equation:
\begin{equation}
\label{eq:1}
\nabla \times \left[({1}/{\varepsilon(\mathbf{r})})\nabla\times \mathbf{H}(\mathbf{r})\right] = ({\omega^2}/{c^2}) \mathbf{H}(\mathbf{r})
\end{equation} 
takes the place of  Schrödinger equation and the localization of electromagnetic resonant modes can be observed for appropriate values of $\beta$.

To provide a relevant framework, we will from now on consider the implementation of the AAH bichromatic lattice exploited in this work, which is schematically shown in Fig.~1 of the main text, superimposed on a scanning electron microscope (SEM) image of one of the fabricated devices. The design starts from a line defect (also known as W1 waveguide) in a triangular lattice of holes \cite{joannopoulos_book}, of period $a$ and radius $r$, patterned in a $220$~nm thick membrane of silicon suspended in air. The W1 is then patterned with a linear lattice of holes, with slightly mismatched periodicity $a'$ and radius $r'<r$, producing a local increase in the effective index of the slab. The secondary lattice is shifted by $\delta=a'/2$ and the mismatch parameter is chosen such that $\beta=N/(N+1)$, with even $N$. The linear lattice of holes recovers then the periodicity of the main PhC after $N+1$ periods of step $a'$, thus yielding a total number of $N+1$ interstitial sites of position $x_j =  j a'$  with $j = 0, \pm 1, \pm 2,\ldots,\pm N/2$. With these parameters, the master equation \eqref{eq:1} describing propagation can be approximated and expressed in the form of the eigenvalues equation:

\begin{equation}
[\omega^2_0 + \Delta \cos(2\pi \beta j)] c_j - J [c_{j-1} + c_{j+1}] = \omega_j^2 c_j 
\end{equation}

where $\omega_0$ is the central frequency of the non-mismatched line defect waveguide modeled in the tight-binding approximation,  $J$ is the tunnel coupling arising from the overlap of basis functions localized in neighboring interstitial sites, and $\Delta$ is the on-site  amplitude of the modulation potential. The eigenvalues $\omega_j^2$ represent the squared resonant frequencies of the localized mode, while the eigenvectors $c_j$ map the distribution of the magnetic field $\mathbf{H}$ on the basis of the Wannier wavefunctions $\mathbf{H}_j$ localized at each interstitial site, via the relation $\mathbf{H} = \sum c_j \mathbf{H}_j$ \cite{Alpeggiani2015}.

\begin{figure*}[t]
\centering
\includegraphics[width=0.8\textwidth]{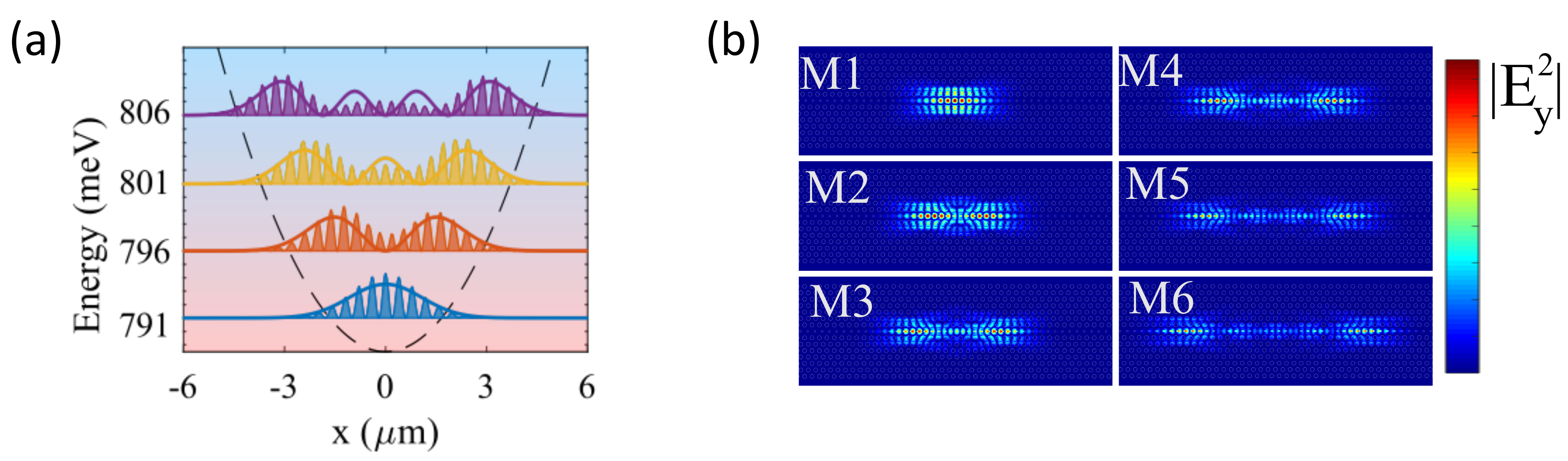}
\caption{(a) Schematic visualization of the AAH confinement mechanism. The effective bichromatic potential can be approximated by a parabola, which yields equally spaced modes with Gauss-Hermite envelope (solid lines). The simulated (FDTD) field distribution of $\left|E_y\right|^2$ ($y=z=0$ slice) finds the best agreement for the lower order modes. For higher order modes, the energy is still concentrated in the side lobes of increasing distance from $x=0~\mu$m. (b) Spatial distribution ($y$-component) of the electric field  $\left|E_y\right|^2$ in the $xy$-plane for the first 6 resonant modes. }\label{fig:supplementary}
\end{figure*}

The effective potential emerging from the AAH exhibits an approximately parabolic shape, analogous to the well-known case of the harmonic oscillator, which in turn results in equally spaced modes \footnote{Strictly speaking, the condition of equal spacing is met for the eigenvalues $\omega_j^2$. However, this distinction is scarcely relevant under the assumption that $\omega_{j+1}-\omega_{j}\ll\omega_0$ } and, remarkably, a Gauss-Hermite profile of the field envelope, as shown in Fig.~\ref{fig:supplementary}a. One of the main consequences is the intrinsic minimization of the out-of-plane losses of the bichromatic cavity modes and thus the natural emergence of high Q factors, as it can be confirmed by looking at the Fourier transform of the electric field distribution in the reciprocal space \cite{Akahane2003}. The value of the mismatch parameter fixes here the convexity of the confinement potential: higher values result in a denser distribution of the modes, higher Q factors and larger mode volumes. The quantities $\Delta$ and $J$, conversely, are linked to the depth of the confinement potential and to the dispersion of the line defect, and are both affected by the design parameter $r'$. It is worth emphasizing that the anharmonicity due to the cosine shape of the bichromatic potential leads to a decreasing trend in the modes spacing for increasing mode order, while the inverse trend (increasing spacing for higher order modes) is provided by the tunneling parameter $J$.

For our design, we fixed $a=400$~nm and $r=100$~nm, in order to achieve a $146$~meV wide photonic bandgap in the telecom band (central energy: $838$~meV), while we chose $\beta\approx0.98$ ($N=48$) and defect hole radii ranging from $r'=50$~nm to $r'=80$~nm, in order to yield a shallow and wide confinement potential. This choice of parameters sets the main difference with our previous work \cite{Simbula2016}, where we focused on the maximization of the figure of merit Q/V. From full three-dimensional finite difference time domain (3D-FDTD) simulations \footnote{All FDTD simulations were performed with a commercial-grade software,\textit{ Lumerical Solutions, Inc}.}, we predicted multi-modal spectra exhibiting more than 8 resonant modes, with an average free spectral range (FSR) of approximately $4$~meV and Q factors higher than $10^7$, as estimated from best fit of each mode ring-down. The mode volumes were estimated to be of the order of $V=2\left(\lambda/n\right)^3$, pointing to the possibility of achieving a large field enhancement effect even for a wide confinement potential profile. A typical (measured) spectrum showing comb-like distribution and the simulated near-field profile for the first 6 modes is shown in Fig.~\ref{fig:supplementary}b, while the $y=0$ slice for the first 4 modes is plotted in Fig.~\ref{fig:supplementary}a, pointing to the analogy with the case of the quantum mechanical harmonic oscillator for the $y$ component of the resonant modes electric field.

% DESIGN AND FABRICATION

\section{Resonant scattering spectroscopy with high absolute accuracy}
\label{sec:rs_technique}

\begin{figure*}[t]
\centering
\includegraphics[width=0.8\textwidth]{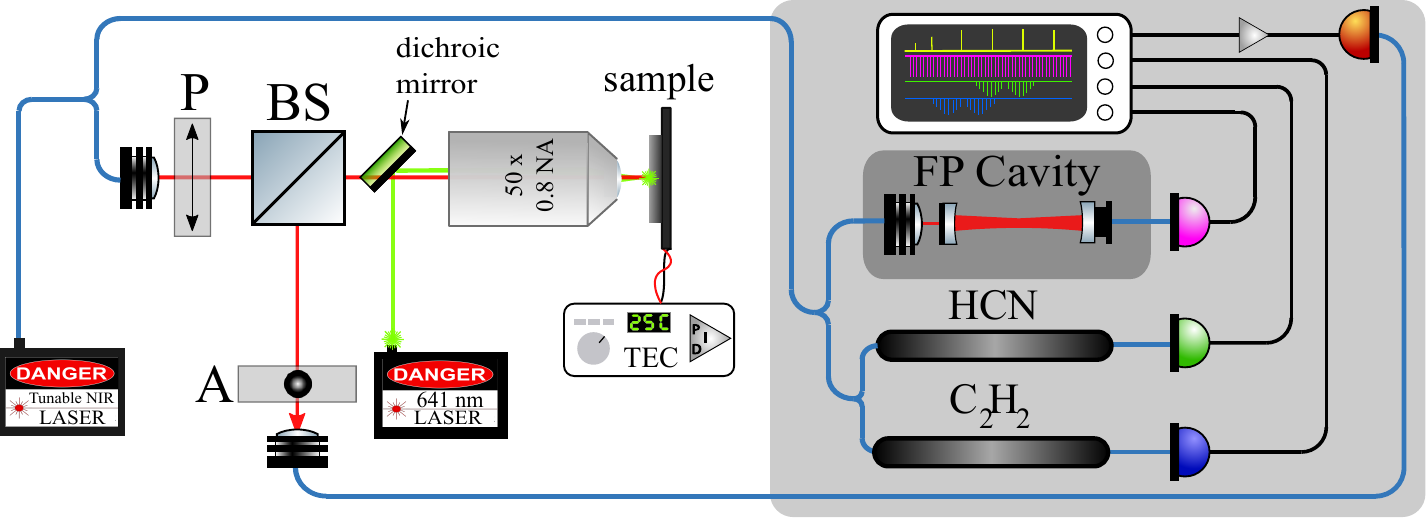}
\caption{Schematic of the experimental apparatus. A tunable laser source operating in the near-infrared (NIR) is swept over the $1500\div1600$~nm band. The radiation is split by fiber beam-splitters (BS) among two reference cells and a Fabry-Pérot (FP) cavity, which combination provides the absolute wavelength reference, and the RS setup. The latter is constituted by a confocal, cross-polarized arrangement where the cavity is placed in the focal plane of a high numerical aperture objective and rotated by $45^\circ$ with respect to the impinging polarization. All the optical signals are collected by single mode optical fibers, photodetected and acquired by a 4-channels oscilloscope. In order to minimize the long-term spectral drift of the resonances, the cavity is stabilized in temperature by means of a thermo-electric cooler (TEC) actuated by a PID controller. The same RS apparatus used for the measurements is exploited for the tuning. A visible laser source is coupled on the same optical path of the IR laser via a dichroic mirror, and focused on the sample with a diffraction-limited spot on the region to be oxidized. }\label{fig:setup}
\end{figure*}

We hereby report the details of the resonant scattering setup employed in this work. A schematic of the experimental apparatus is depicted in Fig.~\ref{fig:setup}. 
The radiation emitted from a tunable external cavity laser (Santec TSL-710) is focused on the sample under examination by means of a confocal microscopy setup. The microcavity is stabilized in temperature ($T_{set}=25\pm0.005~^\circ\mathrm{C}$), in order to compensate for slow thermo-optic drift of the resonant modes, and rotated by $45^\circ$ with respect to the impinging polarization. The crossed polarization is then analyzed and collected by means of a single mode optical fiber, in order to ensure that only the fraction of light resonantly coupled to the optical cavity was effectively observed. The optical signal is then photodetected, amplified and visualized on an oscilloscope triggered by the excitation laser source, which is operated in sweep mode in the $1500\div1600$~nm wavelength range. 

In order to improve the accuracy of our system, every RS spectrum is acquired concurrently with the transmission spectra of two gas reference cells (low-pressure cyanide and acetylene gases) with absorption lines ranging from $1512$~nm to $1564$~nm specified with accuracy of $\pm 0.1$~pm. To further extend the calibrated spectral range, we also measure the transmission through a 1 meter long air-spaced Fabry-Pérot (FP) interferometer ($\mathrm{FSR}_{FP}=0.6~\mu$eV). The acquired spectra is then compared using a linear interpolation algorithm. The combination of the two referencing strategies (gas cells and FP interferometer) enable us to evaluate the resonance frequency of the cavity modes with an overall absolute accuracy of $0.2~\mu$eV, which is below the minimum measured linewidth ($\Gamma_{min}\approx 0.7 ~\mu$eV).

To perform selective oxidation of cavity modes, we employ a visible laser source ($\lambda= 640$ nm), which  is coupled on the same optical path of the IR laser via a dichroic mirror, and focused on the sample with a diffraction-limited spot on the region to be oxidized.

\bibliography{library,library_additional}

\end{document}